\PassOptionsToPackage{table,dvipsnames,usenames}{xcolor}

\documentclass[sigconf]{acmart}

\usepackage{amsmath,amsfonts}
\usepackage{graphicx}
\usepackage{textcomp}
\usepackage[utf8]{inputenc}
\usepackage[linesnumbered,lined,ruled,commentsnumbered]{algorithm2e}
\usepackage{xspace}
\usepackage{multirow}
\usepackage{stfloats}
\usepackage{lscape}
\usepackage{caption}
\usepackage{subcaption}
\usepackage{makecell}
\usepackage{listings, soul} 
\usepackage{colortbl}
\usepackage{booktabs}
\usepackage{courier}
\usepackage{amsthm}
\usepackage{hyperref}
\usepackage{comment}
\usepackage{fontawesome}
\usepackage{pifont}
\usepackage{enumitem}

\definecolor{meta-prompt}{HTML}{2D7D74}  
\definecolor{llm-response}{HTML}{2E7D32} 
\definecolor{light-gray}{gray}{0.95}     

\usepackage{framed}

    \newenvironment{graybox}{%
      \MakeFramed{\advance\hsize-\width\FrameRestore}%
      \raggedright%
    }{%
      \endMakeFramed%
    }

    \newenvironment{whitebox}{%
        \MakeFramed{\advance\hsize-\width\FrameRestore}%
        \raggedright%
    }{%
        \endMakeFramed%
    }

    \newenvironment{promptbox}[1]{%
      \par\medskip%
      \noindent%
      \begingroup%
      \setlength{\fboxsep}{6pt}%
      \colorbox{meta-prompt}{\parbox{\dimexpr\linewidth-12pt}{\color{white}\bfseries#1}}%
      \endgroup%
      \par\nobreak%
      \begin{lrbox}{0}%
        \begin{minipage}{\dimexpr\linewidth-16pt}%
        \vspace{4pt}%
    }{%
        \vspace{4pt}%
        \end{minipage}%
      \end{lrbox}%
      \noindent%
      \setlength{\fboxrule}{1pt}%
      \setlength{\fboxsep}{7pt}%
      \fcolorbox{meta-prompt}{meta-prompt!10!white}{\usebox{0}}%
      \par\medskip%
    }

\setlist[itemize]{noitemsep, topsep=4pt}
\definecolor{lightgray}{HTML}{f6f6f6}
\definecolor{darkgray}{rgb}{.4,.4,.4}
\definecolor{darkblue}{HTML}{1b4db3}
\definecolor{brickred}{HTML}{b04f4f}
\definecolor{purple}{rgb}{0.65, 0.12, 0.82}
\definecolor{diffadd}{HTML}{288f26}
\definecolor{diffrmbg}{HTML}{ffebe9}
\definecolor{diffaddbg}{HTML}{e6ffeb}
\definecolor{diffremove}{HTML}{de4f54}
\definecolor{carrotorange}{rgb}{0.8, 0.33, 0.0}
\definecolor{highlight}{HTML}{fefbc2}

\definecolor{bluegray}{HTML}{3182bd}
\definecolor{delim}{RGB}{20,105,176}
\definecolor{numb}{RGB}{106, 109, 32}
\definecolor{string}{rgb}{0.64,0.08,0.08}

\definecolor{Gray}{gray}{0.6}
\definecolor{LightGray}{gray}{0.9}
\definecolor{bananayellow}{rgb}{1.0, 0.88, 0.5}
\definecolor{myred}{rgb}{1.0,0.44,0.37}

\newcolumntype{A}{>{\columncolor{LightGray}}c}
\newcolumntype{B}{>{\columncolor{LightGray}}r}

\lstdefinelanguage{JavaScript}{
  keywords={typeof, new, true, false, catch, function, return, null, catch, switch, var, const, let, extends, if, in, while, do, else, case, break, async, await, of},
  keywordstyle=\color{darkblue}\bfseries,
  ndkeywords={class, export, boolean, throw, implements, import, this, setTimeout},
  ndkeywordstyle=\color{brickred}\bfseries,
  identifierstyle=\color{black},
  sensitive=false,
  comment=[l]{//},
  morecomment=[f][\color{diffadd}\bfseries]{+\ },
  morecomment=[s]{/*}{*/},
  morecomment=[f][\color{diffremove}\bfseries]{- },
  commentstyle=\color{violet}\ttfamily,
  stringstyle=\color{carrotorange}\ttfamily,
  morestring=[b]',
  morestring=[b]"
}

\lstdefinelanguage{Java}{
	keywords={new, true, false, catch, void, public, return, null, catch, int},
	keywordstyle=\color{darkblue}\bfseries,
	ndkeywords={class, export, boolean, throw, implements, import, this, setTimeout},
	ndkeywordstyle=\color{brickred}\bfseries,
	identifierstyle=\color{black},
	sensitive=false,
	comment=[l]{//},
	morecomment=[f][\color{diffadd}\bfseries]{+\ },
	morecomment=[s]{/*}{*/},
	morecomment=[f][\color{diffremove}\bfseries]{- },
	commentstyle=\color{violet}\ttfamily,
	stringstyle=\color{carrotorange}\ttfamily,
	morestring=[b]',
	morestring=[b]"
}

\lstdefinelanguage{Python}{
	keywords={typeof, new, true, false, catch, function, return, null, catch, switch, var, const, let, extends, if, in, while, do, else, case, break, async, await, of, from, import, class, def},
	keywordstyle=\color{darkblue}\bfseries,
	ndkeywords={class, export, boolean, throw, implements, import, this, setTimeout, self, __init__},
	ndkeywordstyle=\color{brickred}\bfseries,
	identifierstyle=\color{black},
	sensitive=false,
	comment=[l]{//},
	morecomment=[f][\color{diffadd}\bfseries]{+\ },
	morecomment=[s]{/*}{*/},
	commentstyle=\color{violet}\ttfamily,
	stringstyle=\color{carrotorange}\ttfamily,
	morestring=[b]',
	morestring=[b]",
        literate={-}{{{\color{black}-}}}1,
}

\lstdefinestyle{PythonStyle}{
	language=Python,
	backgroundcolor=\color{lightgray},
	extendedchars=true,
	basicstyle=\scriptsize\ttfamily,
	escapeinside={(*@}{@*)},
	showstringspaces=false,
	showspaces=false,
	numbers=left,
	numberstyle=\scriptsize,
	numbersep=6pt,
	tabsize=4,
	breaklines=true,
	showtabs=false,
	captionpos=b,
	frame=single,
	framesep=4pt,
	linewidth=.98\columnwidth,
	xleftmargin=10pt,        
	rulecolor=\color{lightgray},        
        literate={-}{{{\color{black}-}}}1,
}

\lstdefinelanguage{json}{
	numbers=left,
	numberstyle=\scriptsize,
	frame=single,
	rulecolor=\color{lightgray},
	backgroundcolor=\color{lightgray},
	showspaces=false,
	showtabs=false,
	breaklines=true,
	postbreak=\raisebox{0ex}[0ex][0ex]{\ensuremath{\color{gray}\hookrightarrow\space}},
	breakatwhitespace=true,
	basicstyle=\ttfamily\scriptsize,
	upquote=true,
	morestring=[b]",
	literate=
	*{0}{{{\color{numb}0}}}{1}
	{1}{{{\color{numb}1}}}{1}
	{2}{{{\color{numb}2}}}{1}
	{3}{{{\color{numb}3}}}{1}
	{4}{{{\color{numb}4}}}{1}
	{5}{{{\color{numb}5}}}{1}
	{6}{{{\color{numb}6}}}{1}
	{7}{{{\color{numb}7}}}{1}
	{8}{{{\color{numb}8}}}{1}
	{9}{{{\color{numb}9}}}{1}
	{\{}{{{\color{delim}{\{}}}}{1}
	{\}}{{{\color{delim}{\}}}}}{1}
	{[}{{{\color{delim}{[}}}}{1}
	{]}{{{\color{delim}{]}}}}{1},
}

\theoremstyle{definition}

\newcommand{\header}[1]{\par\smallskip\noindent\textbf{#1.}}

\def\BibTeX{{\rm B\kern-.05em{\sc i\kern-.025em b}\kern-.08em
    T\kern-.1667em\lower.7ex\hbox{E}\kern-.125emX}}
    
\newboolean{showcomments}
\setboolean{showcomments}{true}
\ifthenelse{\boolean{showcomments}}
{
	\definecolor{myyellow}{RGB}{255, 228, 26}
	\definecolor{myblue}{RGB}{50, 50, 220}
	\newcommand{\nb}[2]{
		{\sf
			\fcolorbox{myyellow}{yellow}{\scriptsize\textbf{#1}}%
			$\blacktriangleright$%
			{\color{myblue}\fontsize{7pt}{8pt}\selectfont\textbf{#2}}%
		}%
	}
}
{
	\newcommand{\nb}[2]{}
}

\newcommand{\michael}[1]{\nb{Michael}{#1}}

\newcommand{\toolname}{\textsc{\mbox{Issue2Test}}\xspace}

\newcommand{\toolverifiedresolved}{\textsc{30}\xspace}

\newcommand{\toolresolved}{\textsc{84}\xspace}
\newcommand{\toolaccuracy}{\textsc{30.4\%}\xspace}

\newcommand{\code}[1]{{\smaller\ttfamily\texttt{#1}}}

\AtBeginDocument{%
  \providecommand\BibTeX{{%
    Bib\TeX}}}



\begin{document}

\title{
Issue2Test: Generating Reproducing Test Cases from Issue Reports
}

\author{Noor Nashid}
\affiliation{%
\institution{University of British Columbia}
\city{Vancouver}
\country{Canada}}
\email{nashid@ece.ubc.ca}

\author{Islem Bouzenia}
\affiliation{%
\institution{CISPA Helmholtz Center for Information Security}
\city{Stuttgart}	
\country{Germany}}
\email{bouzenia.islem@pm.me}

\author{Michael Pradel}
\affiliation{%
\institution{CISPA Helmholtz Center for Information Security}
\city{Stuttgart}	
\country{Germany}}
\email{michael@binaervarianz.de}    
	
\author{Ali Mesbah}
\affiliation{%
\institution{University of British Columbia}
\city{Vancouver}
\country{Canada}}
\email{amesbah@ece.ubc.ca}

\begin{abstract}
Automated tools for solving GitHub issues are receiving significant attention by both researchers and practitioners, e.g., in the form of foundation models and LLM-based agents prompted with issues.
A crucial step toward successfully solving an issue is creating a test case that accurately reproduces the issue. Such a test case can guide the search for an appropriate patch and help validate whether the patch matches the issue's intent. However, existing techniques for issue reproduction show only moderate success. This paper presents \toolname{}, an LLM-based technique for automatically generating a reproducing test case for a given issue report. Unlike automated regression test generators, which aim at creating passing tests, our approach aims at a test that fails, and that fails specifically for the reason described in the issue. To this end, \toolname{} performs three steps: (1) understand the issue and gather context (e.g., related files and project-specific guidelines) relevant for reproducing it; (2) generate a candidate test case; and (3) iteratively refine the test case based on compilation and runtime feedback until it fails and the failure aligns with the problem described in the issue. We evaluate \toolname{} on the SWT-bench-lite dataset, where it successfully reproduces 32.9\% of the issues, achieving a 16.3\% relative improvement over the best existing technique.
Our evaluation also shows that \toolname{} reproduces 20 issues that four prior techniques fail to address, contributing a total of 60.4\% of all issues reproduced by these tools. 
We envision our approach to contribute to enhancing the overall progress in the important task of automatically solving GitHub issues.
\end{abstract}

	%
	%
	
    
\maketitle

\section{Introduction}
Issue reports, e.g., on GitHub, are commonly used to describe bugs, missing features, and other ways to improve a software project.
Because addressing issues takes significant developer effort, recent work invests heavily in automated issue solving.
Motivated by benchmarks of issues in popular open-source projects, such as SWE-bench~\cite{swebench}, foundation models and Large Language Model (LLM)-based software engineering agents compete in their ability to successfully solve issues~\cite{autocoderover, roy:ai-software-engineer:arxiv25, rondon:agent-based-repair-google:arxiv25, sweagent:arxiv24, openhands:iclr25, coder:arxiv24, magis:neurips25, agentless:arxiv24, pezz:roadmap-se:tosem25}.
In addition to approaches that directly work on issues, automated program repair~\cite{cacm2019-program-repair} tries to address bugs that manifest through a failing test.
Various techniques have been explored for automated bug repair~\cite{fan:code-repair-with-llm:icse23, xia:alpharepair:fse22,repairagent:icse25}.

A crucial prerequisite for successfully solving an issue or repairing a bug is a test case that reproduces the problem.
Unfortunately, issue reports rarely include executable tests that reproduce the issue, making it challenging for developers and automated techniques to validate potential patches.
Specifically in open-source projects, contributors often lack the expertise and familiarity to construct meaningful test cases alongside their reports. 
Most work on automated program repair repair assumes reproducing test cases to be given.
In practice, however, developers often write test cases only after a problem was fixed~\cite{cheng:agentic-bug-reproduction:arxiv25}.
Automated issue solving techniques sometimes include a step for reproducing the issue~\cite{specrover:arxiv24}, but do not explicitly focus on this important part.

One approach for obtaining an issue-reproducing test case could be automated test generation.
Both traditional approaches~\cite{evosuite:fse11, pacheco:randoop:icse07} and neural methods~\cite{atlas:test-assert-generation:icse20, toga:icse22} attempt to reduce the time spent on manually writing tests. However, earlier techniques often fail to generate human-readable and maintainable test cases that align with project conventions and effectively validate intended program behavior~\cite{daka:unit-test-readability:issta17}.
The emergence of LLMs has significantly enhanced automated test generation~\cite{ziegler:copilot-productivity:maps:22, vaithilingam:copilot:chi22, dakhel2022github}. LLMs can synthesize tests with human-like readability and writing style, reflecting patterns learned from extensive training on large code repositories~\cite{ziegler:copilot-productivity:maps:22}. This capability has prompted the use of LLMs in test code generation~\cite{bareiss:test-generation-llm:arxiv22, cedar, lemieux:codamosa:icse23, siddiq:llm-unit-test:arxiv23, schafer:unit-test-generation-using-llm:tse23, xie:chatunitest:arxiv23, software-testing-with-llm-survey}.
While useful for creating regression tests, none of these approaches addresses the problem of creating tests that reproduce a specific issue.

To address the challenge of creating issue-reproducing tests, recent work proposes test generators specifically for this purpose.
Libro~\cite{libro:icse23} generates tests from bug reports using LLM prompting but focuses on structured benchmarks, such as Defects4J, rather than real-world issue reproduction.
Mündler et al.~\cite{niels:bug-fixes-with-code-agents:nips24} introduce SWT-bench-lite, a subset of SWE-bench, for evaluating issue reproduction, and modify the prompts of existing agents for this task.
Yet, these agents are neither designed nor effective at generating issue-reproducing tests.
Most recently, Auto-TDD~\cite{autotdd:arxiv24} proposes a structured multi-step process, achieving a 21.7\% fail-to-pass rate on SWT-bench-lite.
While Auto-TDD improves over prior work, it does not explicitly enforce the creation of failing tests, which is important to successfully reproduce issues.

This paper presents \toolname{}, a novel LLM-based technique for automatically generating a reproducing test case for a given issue report.
In contrast to automated regression test generators, which aim at creating passing tests, our approach aims at a test that fails, and that fails specifically for the reason described in the issue.
To this end, \toolname{} performs three steps:
(1) understand the issue and gather context (e.g., related files and project-specific guidelines) relevant for reproducing it;
(2) generate a candidate test case; and
(3) iteratively refine the test case based on compilation and runtime feedback until it fails and the failure aligns with the problem described in the issue.
Unlike prior work on generating issue-reproducing tests, \toolname{} continuously checks whether the generated tests capture problem described in the issue. 

We evaluate \toolname{} on all 276 instances of SWT-bench-lite, a subset of SWE-bench that has been extensively used for evaluating prior work~\cite{niels:bug-fixes-with-code-agents:nips24,autotdd:arxiv24}. The results show that our approach: i) outperforms all the existing baselines by generating tests for 91 instances (13 more than the best baseline) with the best performing LLM, ii) successfully generates tests for 20 unique issues missed by any prior work and contributes 60.4\% of the union of reproduced issues by any technique, iii) imposes moderate costs of 5.21 cents per issue, which is orders of magnitude cheaper than what a human software developer would charge for performing the same task.

In summary, our paper contributes the following:
\begin{itemize}
    \item The first technique for generated issue-reproducing tests that explicitly aims at producing tests that fail, and that fail for the reasons described in the issue.
    \item Empirical evidence that \toolname{} significantly outperforms the current state of the art, showing a 40.1\% relative improvement over the best existing technique.
    \item We make our code and data publicly available, providing a starting point for future work.
\end{itemize}

\section{Approach}
\label{sec:approach}
In this section, we introduce a real example from the SWT-bench dataset, the challenges accompanying the reproduction of the example issue, how our approach addresses it, step by step, while introducing and explaining the components and the workflow of our approach,  \toolname{}.

\subsection{Running Example}
\label{sec:motivatio}
\definecolor{bggray}{rgb}{1, 1, 1}

The reported Django issue, shown in Figure~\ref{fig:example}, illustrates a regression where overriding \code{get\_FIELD\_display()} no longer works as expected. While the report describes the issue, it lacks essential details for straightforward replication. For instance, the issue is missing details on necessary imports, app declaration, model migrations, and explicit steps to formulate and execute a test case with a failing assertion. Lacking these key elements, developers must dedicate efforts to understanding and filling-up the missing details.
The goal of our paper is to create a technique that suggests ready-to-use test cases that reproduce the problem described by the issue. In the example of Figure~\ref{fig:example}, the issue is a bug, but in other cases it could also reflect a request for a new feature or improved performance.

\begingroup
\setlength{\abovecaptionskip}{2pt}
\begin{figure}
\begin{graybox}
\textbf{Title} 

Cannot override \code{get\_FOO\_display()} in Django 2.2+

\textbf{Description}

I cannot override the \code{get\_FIELD\_display} function on models since version 2.2. It works in version 2.1.

\textbf{Example:}

\begin{lstlisting}[language=Python, frame=single, basicstyle=\ttfamily\footnotesize]
class FooBar(models.Model):
    foo_bar = models.CharField(_("foo"), choices=[(1, 'foo'), (2, 'bar')])

    def __str__(self):
        return self.get_foo_bar_display()  # This returns 'foo' or 'bar' in 2.2, but 'something' in 2.1

    def get_foo_bar_display(self):
        return "something"
\end{lstlisting}

What I expect is that I should be able to override this function.
\end{graybox}
\caption{Django-12284 issue from SWT-bench~\cite{niels:bug-fixes-with-code-agents:nips24}.}
\label{fig:example}
\vspace{-12pt}
\end{figure}
\endgroup

Figure~\ref{fig:gentest} presents a test case generated by \toolname{} for the issue described in Figure~\ref{fig:example}, successfully reproducing the reported failure. Automatically generating such reproducing tests is practically valuable: it helps developers understand the issue more clearly, provides a concrete failure scenario to aid debugging, and enables patch validation in the absence of existing tests. For AI-based tools and agents, the generated test serves as an executable specification of the issue, potentially supporting automated repair and patch validation workflows.


While the running example involves a Django regression, many GitHub issues report failures where the underlying cause must be carefully validated. In such cases, prior approaches often produce tests that fail due to unrelated reasons, such as misconfigured environments or irrelevant assertions, without checking whether the failure corresponds to the reported issue. This underscores the need for verifying that generated test failures are semantically aligned with the issue description.


In this section, we describe the steps and strategies used by our approach to generate such tests.

\definecolor{importcolor}{rgb}{0.3, 0.3, 0.7} 
\definecolor{settingscolor}{rgb}{0.2, 0.6, 0.2} 
\definecolor{helperclasscolor}{rgb}{0.8, 0.4, 0.1} 
\definecolor{testcasecolor}{rgb}{0.7, 0.2, 0.2} 

\begingroup
\setlength{\abovecaptionskip}{2pt}  
\begin{figure}
	\begin{whitebox}
    
        \noindent\textcolor{importcolor}{\textbf{Import Statements}}
		
        
			\begin{lstlisting}[language=Python, basicstyle=\ttfamily\footnotesize]
from django.conf import settings
import django
from django.db import models
from django.test import TestCase
			\end{lstlisting}
		
        \noindent\textcolor{settingscolor}{\textbf{Django Settings Configuration}}
		
        
			\begin{lstlisting}[language=Python, basicstyle=\ttfamily\footnotesize, firstnumber=5]
if not settings.configured:
  settings.configure( INSTALLED_APPS=[
  'django.contrib.contenttypes', 'test_app'],
  DATABASES={'default': {'ENGINE': 
 'django.db.backends.sqlite3','NAME':':memory:'}})
django.setup()\end{lstlisting}
		
        \noindent\textcolor{helperclasscolor}{\textbf{Target Test Subjects (Models A and B)}}
		
			\begin{lstlisting}[language=Python, basicstyle=\ttfamily\footnotesize, firstnumber=11]
class A(models.Model):
  foo_choice = [("A", "output1"),("B", "output2")]
  field_foo = models.CharField(max_length=254, 
                    choices=foo_choice)
  class Meta:
    abstract = True

class B(A):
  foo_choice = [("A", "output1"), ("B", "output2")
                ,("C", "output3")]
  field_foo = models.CharField(max_length=254,
                    choices=foo_choice)
  class Meta:
    app_label = 'test_app'
			\end{lstlisting}
		
        \noindent\textcolor{testcasecolor}{\textbf{Test Case and Assertion}}
		
        
			\begin{lstlisting}[language=Python, basicstyle=\ttfamily\footnotesize, firstnumber=25]
class ModelDisplayTests(TestCase):
  def test_get_field_foo_display_for_C(self):
    instance = B(field_foo='C')
    self.assertEqual(instance.get_field_foo_display(),'output3')
			\end{lstlisting}
    \end{whitebox}
\caption{Test generated by \toolname{} for Django-12284.}
\label{fig:gentest}
\end{figure}
\endgroup

\begin{figure*}
    \centering
    \includegraphics[width=1\linewidth]{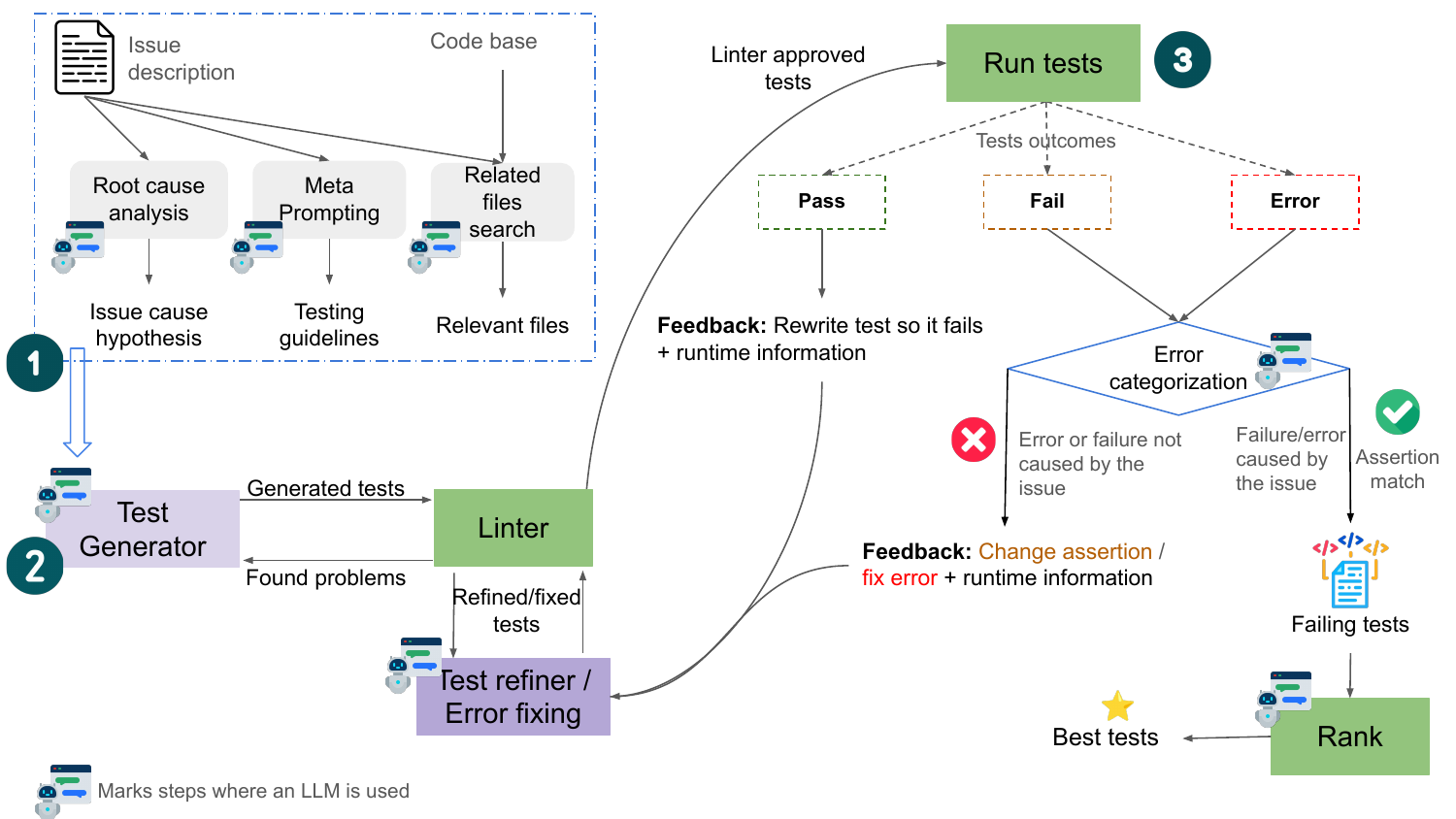}
    \caption{Approach overview}
    \label{fig:architecture}
\end{figure*}

\subsection{Overall Design}
The overall design of our approach is shown in Figure \ref{fig:architecture}.
\toolname{} automates the task of test generation by combining structured LLM prompting, validation based on runtime feedback, and iterative refinement. Our technique provides a pipeline for generating and refining test cases derived from reported issues (e.g, the issue in Figure~\ref{fig:example}).
The approach is structured into three main phases, briefly described in the following and then in detail in the subsequent subsections.
Our design invokes LLMs throughout various steps of the approach, as shown in Figure~\ref{fig:architecture}.

Phase~1, \emph{issue comprehension}, analyzes the root cause of the problem, infers project-specific testing guidelines, and identifies relevant files.
\toolname{} begins by analyzing a reported issue to extract essential context, including the nature of the problem, its potential causes, and the associated code and test files. During the first phase, the tool queries an LLM to conduct an initial root cause analysis, yielding a list of plausible reasons. Subsequently, a set of project-specific guidelines are generated that encapsulate best practices and conventions for addressing issues in the target project, along with recommendations for writing isolated standalone test cases. The final step in this phase involves searching the codebase to locate relevant files based on the project’s structure.

Phase~2, \emph{test generation}, synthesizes an initial test case.
The information gathered from Phase~1, combined with the issue details, is used to construct a test generation prompt. \toolname{} then queries an LLM with this prompt and expects a Python script containing the testing code.
The generated script is then passed through a static linter to detect any potential issues. If the linter identifies problems, the test generation prompt is augmented with the linter's feedback, and the LLM is subsequently asked to resolve these issues.

Phase~3, \emph{test refinement}, iteratively improves the tests based on execution feedback until it fails for the reasons described in the issue.
To this end, the approach executes the generated test case. Each test case can result in one of three outcomes: pass, fail, or error. For test cases that pass, \toolname{} queries the LLM to modify the test behavior so that it appropriately fails in response to the described issue. For test cases ending in failure or an error, \toolname{} first determines whether the failure is related to the reported issue; if not, the LLM is prompted to refine the assertion. Conversely, if the failure is relevant (either a test assertion or a runtime error), the test case is added to the set of candidate tests.
Finally, \toolname{} ranks all candidate tests based on their relevance and outputs a single test case.

\subsection{Phase 1: Issue Comprehension}  
The goal of this phase is to collect contextual and informative ingredients to construct the \textit{test generation prompt} used in the subsequent phase.

\header{Root Cause Analysis} A typical first step, in creating tests for an issue, is to analyze its description and extract possible causes. This simple analysis allows to have an initial hypothesis about the reasons behind the issue and get a first idea on what behvaior should be reproduced and how. Capitalizing on this simple principle, \toolname{} systematically identifies root causes by prompting an LLM to do the root cause analysis.
As shown in Figure~\ref{fig:root-cause-prompt}, we prompt the LLM to suggest different possible sets of reasons that could be behind the issue. As one issue might have multiple plausible reasons,  the LLM is also asked to suggest sets of \textit{mutually exclusive causes} where validating one negates another. For example, an issue may arise from an incorrect API call or an outdated dependency—distinct causes requiring separate resolutions. In Figure~\ref{fig:rootcauseresult}, we show the resulting analysis for Django-12284. The third cause suspected by the LLM is the correct one and the one that led into generating a succesful reproduction of the issue shown in Figure~\ref{fig:gentest}. In our approach, each cause is used in a different test generation prompt later on. That is, in our running example, \toolname{} would construct three independent test generation prompts, each assuming one of the three causes.

\begingroup
\setlength{\abovecaptionskip}{2pt}  
\begin{figure}[htbp]
    \centering
    \begin{promptbox}{Root Cause Analysis Prompt}
    \textbf{Task:} Perform root cause analysis for the following GitHub issue: Cannot override \code{get\_FOO\_display()} in Django 2.2+ Description ...(shortened for paper)

    \begin{itemize}[left=0pt]
        \item Propose distinct sets of root causes that could explain the issue.
        \item Integrate contextual information from the issue description. 
        \item Identify mutually exclusive or independent failure scenarios.
    \end{itemize}
    \end{promptbox}

    \caption{Root cause analysis prompt for Django-12284.}
    \label{fig:root-cause-prompt}
\end{figure}
\endgroup

\begingroup
\setlength{\abovecaptionskip}{2pt}
\begin{figure}
\begin{graybox}

\textbf{Cause 1: Handling of Choices in Inherited Models} 

... (shortened for the paper)

\textbf{Cause 2: Cache or Meta-Class Processing}

... (shortened for the paper)

\textbf{Cause 3: Implementation Bug in Django's Display Method} 

\textit{--> Reason 1:} There might be a bug within Django itself where the \code{get\_FOO\_display()} method does not account for inherited or overridden choices correctly, particularly when choices are altered at the child class level.

\textit{--> Reason 2:} The method that resolves choices within Django might be erroneously falling back to a default or initial state rather than considering the dynamic changes made by child classes, leading to incorrect value displays.

\end{graybox}

\caption{Root cause analysis result for Django-12284.}
\label{fig:rootcauseresult}
\end{figure}
\endgroup

\header{Project-Specific Guideline Generation} The automated test generation problem requires knowledge of the target project's testing frameworks, conventions, and dependency structures. A general approach is insufficient and cannot cover all cases, as different projects impose distinct constraints even compared to their own different versions. For instance, Django requires specific fixture configurations for test execution. Ideally, developers writing unit tests are familiar with a project's structure and setup, allowing them to adhere to established conventions. However, manually crafting repository-specific test guidelines for each project is impractical due to the time and effort required.

Inspired by ExecutionAgent~\cite{agent-test-execution:arxiv24}, we employ the principle of \textit{meta-prompting} which enables the LLM to dynamically retrieve structured guidelines tailored to the repository before generation. Meta-prompting serves as an adaptive knowledge retrieval mechanism, leveraging the LLM’s pre-trained knowledge to synthesize project-specific testing conventions without relying on predefined templates. For instance, we send a query to the LLM, incorporating the repository name, version, and issue description alongside instructions to generate concise guidelines for writing executable unit tests.
The LLM's response is then used in further prompts to provide project-specific context for the test generation task.

\begingroup
\setlength{\abovecaptionskip}{2pt}  
\begin{figure}[t]
    \centering
    \begin{promptbox}{Guidelines Meta Prompt}
    \textbf{Issue:} 
    
    [Excerpts from Django-12284, shortened for the paper]
    
    Cannot override get\_FOO\_display() in Django 2.2+ ...

    \textbf{Task:} Provide structured guidelines for writing a self-contained and executable unit test for the issue above, which was reported in the project \texttt{Django 2.2+}.
    \end{promptbox}

    \caption{Generating project-specific test guidelines for Django-12284}
    \label{fig:meta-prompt}
\end{figure}
\endgroup

Figure~\ref{fig:meta-prompt} illustrates a meta-prompt example for Django-12284, where \toolname{} queries the LLM to derive testing guidelines specific to the Django project and version. The prompt explicitly provides the GitHub issue description, along with the project name (\texttt{Django}) and version (\texttt{2.2+}), instructing the LLM to extract relevant testing conventions, such as required configurations, fixture setups, and execution constraints.
LLMs have been trained on a vast number of open-source projects, allowing them to capture diverse project-specific practices and requirements, which can provide useful insights when working with similar repositories. However, the knowledge they utilize in their responses depends heavily on how the prompt is structured. Providing clear and detailed instructions in the prompt helps ensure more relevant and accurate outputs.
By integrating meta-prompting, we aim to eliminate the need for manually curated test-writing instructions while ensuring adherence to project conventions.

\header{Locating Related Files} Developers analyze the repository structure, source code, and test files to identify those relevant to a reported issue. Similarly, we generate a structured repository representation that enables the LLM to infer relevant files. Instead of scanning the entire repository, we construct a hierarchical tree format that preserves directory relationships, test files, and source code listings. A recursive traversal extracts directory names and file paths while maintaining structural depth. This structured representation, provided alongside the reported issue, allows the LLM to rank the most relevant files. For the running example, the list of related files is shown in Figure~\ref{fig:retrieved-files}.

\begingroup
\setlength{\abovecaptionskip}{2pt}
\begin{figure}[t]

    \begin{graybox}

\textbf{IssueID:} django-12284

\textbf{Found files:}

1. django/db/models/options.py

2. django/db/models/fields/\_\_init\_\_.py

3. django/db/models/fields/mixins.py

4. django/db/models/base.py

5. django/db/models/query.py

   \end{graybox}
    \caption{Example of retrieved related files for Django-12284.}
    \label{fig:retrieved-files}
\end{figure}
\endgroup

\subsection{Phase 2: Test Generation}

As shown in Figure~\ref{fig:architecture}, we use the information from the previous step to construct the test generation prompt. The goal of this prompt is to instruct the model to generate test cases that fail, so that the tests capture the behavior described by the issue. In Figure~\ref{fig:structured-prompt}, we provide an example of a test generation prompt for Django-12284.
Using the constructed prompt, the approach queries the LLM to generate a test script, extracts the Python code from the response, and writes it into a file.

\begingroup
\setlength{\abovecaptionskip}{2pt}  
\begin{figure}[t]
    \centering
    \begin{promptbox}{Test Generation Prompt}
\textbf{Task:}
Your task is to generate unit tests based on the provided GitHub issue description and source code. Each test case should be designed to fail initially, confirming the existence of a bug or unimplemented feature.

\textbf{GitHub Issue:}
Cannot override get\_FOO\_display() in Django 2.2+ 
Description ...(shortened for brevity)

\textbf{Guidelines:}
To create an independent, executable unit test in Django, dynamically configure settings and use an in-memory database for isolation.
Configure Django Settings...(shortened)

\textbf{Root cause analysis:}
The Django model's mechanism for resolving choices in an inherited class might not correctly override ...

\textbf{Related Source File:}
\begin{lstlisting}[language=Python, frame=single, basicstyle=\ttfamily\footnotesize]
import collections.abc
import copy
... (shortened for paper)
def rel_db_type(self, connection):
  return SmallIntegerField().
         db_type(connection=connection)
\end{lstlisting}

\textbf{Related Test File:}
\begin{lstlisting}[language=Python, frame=single, basicstyle=\ttfamily\footnotesize]
... (shortened for paper)
class BasicFieldTests(SimpleTestCase): 
  def test_show_hidden_initial(self):
... (shortened for paper)
  self.assertChoicesEqual(
  field.get_choices(include_blank=False, limit_choices_to={}),[self.bar1, self.bar2],)
\end{lstlisting}

\textbf{Expected Output:}
Self-contained Python test file.
    \end{promptbox}
    \caption{Test Generation Prompt for Django-12284}
    \label{fig:structured-prompt}
\end{figure}
\endgroup


\begingroup
\setlength{\abovecaptionskip}{2pt}  
\begin{figure}[t]
    \centering
    \begin{promptbox}{Error Categorization Prompt}
\textbf{Task:}

Analyze the outcome of the below-failing test and classify it into one of the following:

1. Compilation Error: The test cannot be compiled.

2. Runtime Error: The test crashes during execution.

3. Assertion Failure: The test runs and fails due to an assertion.

Also, tell whether the assertions or error is caused by the issue described below.

\textbf{GitHub Issue:}
Cannot override \code{get\_FOO\_display()} in Django 2.2+ Description ...(shortened for paper)

\textbf{Test Execution Output:}
\begin{lstlisting}[language=Python, frame=single, basicstyle=\ttfamily\footnotesize, numbers=none]
FAIL: test_get_field_foo_display_for_C 
Traceback (most recent call last):
  File "/testbed/tests/model_fields/test_new.py", line 43, in test_get_field_foo_display_for_C   self.assertEqual(instance.get_field_foo_display(), 'output3')
AssertionError: 'C' != 'output3'
- C
+ output3
\end{lstlisting}

\textbf{Expected Response:}
\begin{lstlisting}[language=json, frame=single, basicstyle=\ttfamily\footnotesize, numbers=none]
{
  "error_type": "<compilation, runtime, or assertion>",
  "issue_error_relevance": <true or false>,
  "repair_steps": "<corrective actions>"
}
\end{lstlisting}
    \end{promptbox}
    
    \caption{Error categorization for Django-12284}
    \label{fig:error-categorization}
\end{figure}
\endgroup

\subsection{Phase 3: Test Refinement} 
The third phase is the core contribution of \toolname{}, as it specifically aims at producing a test that fails, and that the failure is due to the reason described in the issue.
To this end, the test code generated in the previous step is executed within a Docker environment to validate its behavior. The generated test is placed into a new file in the project's test suite and executed in an isolated container to prevent external dependencies from influencing the results. After test execution, logs are generated to capture the test output. These logs are then analyzed to verify the outcome of the test execution.

Next, \toolname{} iteratively refines test cases to ensure they accurately capture the reported issue. It operates in a feedback-guided manner, where test execution outcomes influence subsequent steps. The process is structured around two main objectives: (i) generating an initial failing state before the issue is resolved and (ii) addressing failures unrelated to the reported issue. The high-level refinement process is shown in Algorithm~\ref{alg:test-refinement}.  

Algorithm~\ref{alg:test-refinement} begins by running the generated test (Line~\ref{alg1-1}) and analyzing the results. If a test failure occurs, \emph{error categorization} (Line~\ref{alg1-3}) determines the failure type and whether a \emph{runtime error is directly related to the reported issue}. If a runtime failure is due to the issue, the refinement process terminates successfully (Line~\ref{alg1-4}). The corresponding prompt is shown in Figure~\ref{fig:error-categorization}, guiding this classification.  

Next, for assertion failures, \toolname{} checks whether the failure directly corresponds to the GitHub issue (Line~\ref{alg1-5}). If the assertion failure is issue-related, the test is considered valid. Otherwise, the test is further refined to better align with the issue.
If no failures occur or if the failure is unrelated to the issue, a refinement step is applied (Line~\ref{alg1-2}). 
The refinement step aims at transforming the currently passing test into a failing test, e.g., by adjusting the  assertions.   
The prompt for this transformation is shown in Figure~\ref{fig:modify-passing-test-prompt} for Django-12284, where \toolname{} instructs the LLM to modify an initially passing test to ensure it fails due to the reported issue. If an assertion failure occurs but is unrelated to the issue, \toolname{} prompts the LLM to refine the assertion, ensuring that it directly captures the expected behavior associated with the reported defect.

For compilation and runtime failures that are not directly related to the issue, \toolname{} applies targeted fixes (Line~\ref{alg1-6}). Specifically, it searches the repository for missing imports and incorporates this information into the prompt to guide the LLM in refining the test case.

In the final step, outside the feedback loop, the approach selects the best test case among all failing tests produced so far.
The previous three phases may produce multiple tests, e.g., by working under different hypotheses regarding the root cause of the problem.
To select the best test, we prompt the LLM with the GitHub issue, asking it to rank the generated test cases and retain only those relevant to the issue. 

\begingroup
\setlength{\abovecaptionskip}{2pt}  
\begin{figure}[t]
    \centering
    \begin{promptbox}{Test Modification Prompt(pass into fail)}
\textbf{Task:}

The test case below passes even though the reported GitHub issue is unresolved. Modify the test case to ensure that it fails because of the issue below.

\textbf{GitHub Issue:}

Cannot override \code{get\_FOO\_display()} in Django 2.2+ Description
...(shortened for paper)

\textbf{Passing Test:}

\begin{lstlisting}[language=Python, frame=single, basicstyle=\ttfamily\footnotesize]
from django.conf import settings
import django
... (shortened for paper)
  def test_get_field_foo_display_for_A(self):
    instance = B(field_foo='A')
    self.assertEqual(instance.get_field_foo_display(), 'output1')

\end{lstlisting}

    \end{promptbox}
    
    \caption{Prompt used to transform a generated, passing test into a failing one.}
    \label{fig:modify-passing-test-prompt}
\end{figure}
\endgroup

\begin{algorithm}[t]
    \SetAlgoLined
    \DontPrintSemicolon
    \caption{Execution-Driven Test Refinement}
    \label{alg:test-refinement}
    \footnotesize{
    \KwIn{Test case $T_0$, GitHub issue $G$, Source Code $S$}
    \KwOut{Refined failing test case $T^*$ or failure}

    $T \gets T_0$, $iter \gets 0$, $maxIterations \gets 20$\;

    \While{$iter < maxIterations$}{
        $testResults \gets \textbf{RunTestAndCaptureErrors}(T)$\;
        $passingTests, failingTests \gets \textbf{AnalyzeTestResults}(testResults)$\;\label{alg1-1}

        \tcp{Step 1: Categorize Failure and Identify Runtime Relevance}
        $(failureType, isRuntimeFailureRelated) \gets \textbf{error\textunderscore categorization}(failingTests, G)$\;\label{alg1-3}

        \tcp{Step 2: Stop If Runtime Failure is Issue-Related}
        \If{$isRuntimeFailureRelated$}{\label{alg1-4}
            \Return $T$ \tcp*[r]{Runtime failure correctly captures issue}
        }

        \tcp{Step 3: Handle Assertion Failures}  
        $isAssertionFailureRelated \gets \textbf{false}$
        
        \If{$failureType = \text{assertion}$}{\label{alg1-5}
            $isAssertionFailureRelated \gets \textbf{assertion\textunderscore match}(failingTests, G)$\;
            \If{ $isAssertionFailureRelated$}{
                \Return $T$ \tcp*[r]{Assertion failure correctly reflects issue}
            }
        }

        \tcp{Step 4: Refine Test to Fail and Avoid Unrelated Assertion Failures}
        \If{($failureType = \text{passing}$) \textbf{ or } (\textbf{not} $isAssertionFailureRelated$)}{
            $T \gets \textbf{test\textunderscore refinement}(T, G)$\;\label{alg1-2}
            $iter \gets iter + 1$\;
            \textbf{continue}; \tcp*[r]{Retry with refined test}
        }

        \tcp{Step 5: Handle Compilation or Runtime Failures (Only if no prior modification)}~\label{alg1-6} 
        \If{$failureType = \text{compilation}$ \textbf{or} $failureType = \text{runtime}$}{
            $T \gets \textbf{runtime\textunderscore compilation\textunderscore fix}(T, failingTests, G, S)$
        }
        $iter \gets iter + 1$
    }
    \Return $\emptyset$\;
    }
\end{algorithm}

\section{Evaluation}
\label{sec:evaluation}
To assess the effectiveness of \toolname{} we address the following research questions: 

\newlist{researchquestions}{enumerate}{1}
\setlist[researchquestions]{label*=\textbf{RQ\arabic*}}

\begin{researchquestions}    
    \item What is the effectiveness of \toolname{} in generating fail-to-pass tests, and how does it compare to the state-of-the-art?
    
    \item How do different components of the approach influence the test generation results?

    \item What are the token consumption and LLM invocation costs of \toolname{}?
\end{researchquestions}

\subsection{Experimental Setup}

\subsubsection{Benchmark} 
We use the SWT-bench-lite benchmark, proposed recently by Mündler et al.~\cite{niels:bug-fixes-with-code-agents:nips24}, as the benchmark for our evaluation. SWT-bench-lite is a curated subset of the SWT-bench dataset, designed to cost-efficiently evaluate the ability of language models to generate unit tests that reproduce real-world software issues.
It consists of 276 instances derived from popular GitHub repositories, each containing a GitHub issue description, a corresponding issue-resolving patch, and a set of golden reference test cases that fail on the original version but pass after the issue has been resolved. Pull requests included in SWT-bench-lite meet the following criteria: (i) they are merged into the main branch, (ii) they explicitly link to a resolved GitHub issue, and (iii) they modify at least one test file. 

In addition to the full SWT-bench-lite dataset (called \emph{all issues}), we also report results on those 90 instances from SWT-bench-lite that are part of SWE-bench Verified~\cite{swebench-verified} (called \emph{verified issues}).
SWE-bench Verified~\cite{swebench-verified} is a human-validated subset of the original SWE-bench~\cite{swebench}. The validation process for SWE-bench Verified, conducted by OpenAI, involved expert annotation to ensure that each GitHub issue is solvable, clear, and linked to a valid bug-fixing patch. Each instance was manually reviewed to confirm that the issue description provides sufficient context and that the patch and test cases accurately capture the issue and its resolution. 

\subsubsection{LLM Empowering \toolname{}}
For all our experiments, we employ GPT-4o-mini (version 2024-07-18), a cost-efficient proprietary model from OpenAI, as our primary model for evaluation. GPT-4o-mini is smaller than GPT-4o but faster, offering a 128K context window. At the time of writing the paper, the used model costs \$0.15 per million input tokens and \$0.60 per million output tokens. 

To assess the impact of different models, we further evaluate \toolname{} with Meta Llama 3.3-70B (version meta.llama3-3-70b-instruct-v1:0 from AWS Bedrock)~\cite{llama}, an open-source model, and Claude 3.5 Sonnet (version 20241022)~\cite{claude}, a leading proprietary model from Anthropic.

\subsubsection{Evaluation Metrics}
\label{subsec:evaluation-metrics}
To evaluate the effectiveness of generated test cases, we use the \emph{fail-to-pass (F→P) rate}, as used in prior work~\cite{niels:bug-fixes-with-code-agents:nips24,autotdd:arxiv24}.
Given one test generated per issue, the metric measures the proportion of generated tests that are issue-reproducing and patch-validating, i.e., they initially fail on the pre-patch code and pass after the issue-resolving patch is applied. A high F→P rate indicates that the generated tests correctly reproduce the issue and validate its resolution, making it a crucial measure of test generation quality.

\subsubsection{Baselines}
\label{sec:baselines}
To evaluate \toolname{}, we compare it against several baselines that represent different approaches to test generation, following prior work by Mündler et al.~\cite{niels:bug-fixes-with-code-agents:nips24} and Auto-TDD~\cite{autotdd:arxiv24}. Zero-Shot Prompting (ZeroShot) uses direct LLM prompting to generate test cases based on issue descriptions. Libro~\cite{libro:icse23} was the first approach specifically designed for generating bug-reproducing test cases. Code agents SWE-Agent~\cite{sweagent:arxiv24} and AutoCodeRover~\cite{autocoderover} were originally developed for issue solving, but were adapted for test generation using a prompt to create issue-reproducing unit tests~\cite{niels:bug-fixes-with-code-agents:nips24}. SWE-Agent+~\cite{niels:bug-fixes-with-code-agents:nips24} extends SWE-Agent by executing the generated tests before finalizing them, improving issue-specific test generation. Finally, Auto-TDD~\cite{autotdd:arxiv24} is a recent approach focusing on test generation for GitHub issues similar to \toolname{}.

To compare \toolname{} with publicly available techniques, we include non-peer-reviewed entries from the SWT-Bench Lite leaderboard\footnote{\url{https://swtbench.com}}.
This results in the inclusion of two variants of OpenHands~\cite{openhands:iclr25}, a generic-purpose coding agent. The first variant, OpenHands-Vanilla, uses the agent as released; the second variant, OpenHands-CI, augments the agent with a test‐code‐specific feedback extraction mechanism.

\subsection{Effectiveness and Comparison (RQ1)} 

\subsubsection{\toolname{}'s effectiveness}

\begin{table}
    \caption{\toolname{} results per project on SWT-bench-lite.}
    \label{tab:swt_summary}
    \centering
    \small
    \setlength\tabcolsep{4.5pt}
    \begin{tabular}{@{}l|rrr|rrr@{}} \toprule
        \multirow{2}{*}{\bf Project} & \multicolumn{3}{c}{\bf Verified issues} & \multicolumn{3}{c}{\bf All issues} \\  
        \cmidrule(lr){2-4} \cmidrule(lr){5-7}
        & \bf Issues & \bf F→P & \bf F→P (\%) & \bf Issues & \bf F→P & \bf F→P (\%) \\  
        \midrule
        django               &   42 &    8 &  19.05 &  113 &   18 &  15.93 \\
        sympy                &   22 &   11 &  50.00 &   71 &   34 &  47.89 \\
        scikit-learn         &    9 &    5 &  55.56 &   23 &   13 &  56.52 \\
        matplotlib           &    7 &    2 &  28.57 &   23 &    7 &  30.43 \\
        astropy              &    4 &    3 &  75.00 &    6 &    3 &  50.00 \\
        sphinx-doc           &    3 &    0 &   0.00 &   11 &    0 &   0.00 \\
        pydata               &    1 &    1 & 100.00 &    5 &    2 &  40.00 \\
        pytest-dev           &    1 &    0 &   0.00 &   11 &    3 &  27.27 \\
        pylint-dev           &    1 &    0 &   0.00 &    6 &    3 &  50.00 \\
        pallets              &    0 &    0 &   0.00 &    2 &    0 &   0.00 \\
        psf                  &    0 &    0 &   0.00 &    1 &    1 & 100.00 \\
        mwaskom              &    0 &    0 &   0.00 &    4 &    0 &   0.00 \\
        \midrule
        \textbf{Total/Avg} &   90 &   30 &  33.33 &  276 &   84 &  30.43 \\
        \bottomrule
    \end{tabular}
    \vspace{-10pt}
\end{table}

Table~\ref{tab:swt_summary} presents the effectiveness of \toolname{} in generating F→P test cases per project using GPT-4o-mini. The results are reported separately for verified issues and all issues in the SWT-bench-lite dataset.
Across all projects, \toolname{} achieves an overall F→P rate of \toolaccuracy, successfully generating F→P tests for \toolresolved out of 276 issues. On the verified subset, \toolname{} reproduces \toolverifiedresolved out of 90 cases.  

The results vary across projects. \toolname{} performs best on \emph{astropy}, achieving 75.0\% F→P rate on the verified issues and 50.0\% overall, though this project has relatively few issues. \emph{Scikit-learn} also shows strong performance, with 55.6\% F→P rate on verified issues and 56.5\% overall. \emph{Sympy} maintains a rate close to 50\% in both categories. In contrast, \toolname{} cannot generate any F→P test on the verified issues of \emph{sphinx-doc} and \emph{pytest-dev}, indicating challenges in generating effective test cases for these projects.

\subsubsection{Comparison with Baselines}


\begin{table}[t]
    \caption{Comparing \toolname{} with baseline approaches.}
    \label{tab:fail_to_pass}
    \centering
    \begin{tabular}{l rr}
        \hline
        \textbf{Approach} & \multicolumn{2}{c}{\textbf{F→P}} \\
        \cmidrule{2-3}
        & \textbf{Total} & \textbf{Rate} \\
        \hline
        ZeroShot & 16 & 5.8\% \\

        AutoCodeRover~\cite{autocoderover} & 25 & 9.1\% \\
        
        ZeroShotPlus & 28 & 10.1\% \\
        
        Libro~\cite{libro:icse23} & 42 & 15.2\% \\        

        SWE-Agent & 46 & 16.7\% \\
        SWE-Agent+ & 53 & 19.2\% \\

        Auto-TDD~\cite{autotdd:arxiv24} & 60 & 21.7\% \\

        OpenHands-Vanilla (Claude 3-5 Sonnet) & 63 & 22.8\% \\
        OpenHands-CI (Claude 3-5 Sonnet) & 78 & 28.3\% \\
        
        \hline
        \toolname{} (Llama 3.3) & 49 & 17.8\% \\
        \textbf{\toolname{} (GPT-4o-mini)} & \textbf{\toolresolved} & \textbf{\toolaccuracy} \\
        \textbf{\toolname{} (Claude 3-5 Sonnet)} & \textbf{91} & \textbf{32.9\%} \\
        \hline
    \end{tabular}
\end{table}

Table~\ref{tab:fail_to_pass} compares \toolname{} against the baselines (Section \ref{sec:baselines}) 
using the same 276 instances from SWT-bench-lite.
Among the baselines, ZeroShot achieves the lowest F→P rate, being successful for only 5.8\% of the issues. AutoCodeRover and Libro perform better, achieving 9.1\% and 15.2\%, respectively. ZeroShotPlus improves upon ZeroShot by leveraging a custom diff format designed to be more robust for LLM-generated patches. It is worth noting that while AutoCodeRover and SWE-Agent are primarily designed for resolving GitHub issues through prompt modification, they are still capable of generating a moderate number of patch-validating tests.
SWE-Agent outperforms these techniques with a 16.7\% F→P rate, further improving to 19.2\% with SWE-Agent+, which incorporates execution-based filtering. The improvement in SWE-Agent+ comes from the incorporation of execution-based filtering, where generated tests are executed on the pre-patch code, and those that do not fail as expected are discarded. However, no explicit refinement is performed to correct failing tests that do not fully capture the reported issue, limiting its ability to handle assertion mismatches or refine test logic for more precise issue reproduction. Auto-TDD achieved the highest F→P rate among prior methods, generating F→P tests for 21.7\% of issues. \toolname{} outperforms Auto-TDD, achieving {\toolaccuracy} and resolving {\toolresolved} issues. 

OpenHands-Vanilla achieves 22.8\% F→P rate, demonstrating that a general-purpose agent without project-specific setup cannot extract test execution feedback during test generation. 
In contrast, OpenHands-CI is augmented with a continuous-integration pipeline: it installs all dependencies from the project manifest, injects the canonical test-suite invocation, and discards any diffs outside the existing tests/ directory. After each generated patch, the CI harness reruns the full suite and returns the failure trace. This structured feedback raises the F→P rate to 28.3 \%, an improvement over the vanilla configuration. These results suggest that generic-purpose agents require task-specific adaptation for effective automated test generation.

Among the three LLMs employed in \toolname{}, Llama 3.3 yields the lowest F$\!\rightarrow$P accuracy. GPT-4o-mini improves the accuracy to 30.4\% at \$0.0521 per issue, offering the best cost–performance balance. \toolname{} with Claude 3.5 Sonnet attains the highest accuracy—32.9\%; however, it requires \$0.66 per issue, a cost substantially higher than GPT-4o-mini. Unless otherwise stated, the remainder of this paper presents results obtained with the GPT-4o-mini configuration.

Overall, \toolname{} differs from existing techniques by combining test validation with iterative refinement, resulting in more accurate and relevant reproducing tests. It analyzes whether failures, such as compilation errors, runtime exceptions, or assertion mismatches, align with the issue description, rather than treating any failure as sufficient. Diagnostic signals, including stack traces and assertion errors, are incorporated into a feedback loop that guides the LLM toward reproducing the issue. Furthermore, \toolname{} tailors test generation to project-specific conventions through meta-prompting, which is critical in frameworks where improper test scaffolding causes runtime errors, and it generates different test cases for mutually exclusive root-cause hypotheses, each reflecting a plausible explanation for the issue.

\begin{figure}
    \centering
    \includegraphics[width=0.80\linewidth]{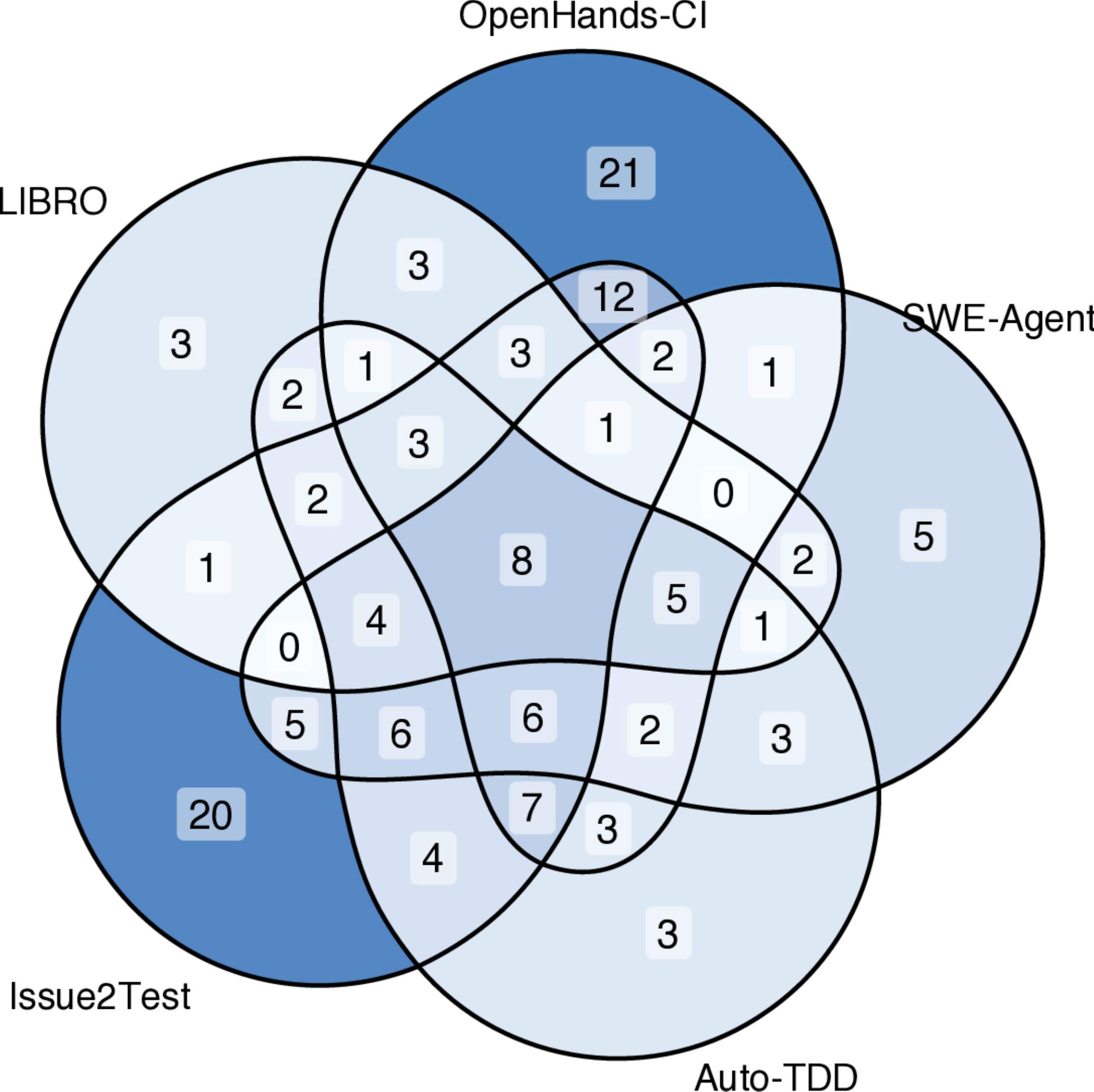}
    \caption{Venn diagram of F→P tests across approaches.}
    \label{fig:venn-diagram}
\end{figure}

Figure~\ref{fig:venn-diagram} presents a Venn diagram comparing the F→P tests generated by \toolname{} and four baseline techniques, including Libro, SWE-Agent+, Auto-TDD and OpenHands-CI. The diagram shows both unique and overlapping F→P tests among these methods.
%
\toolname{} generates 20 unique F→P tests not generated by any other approach.
This indicates that \toolname{} can generate a significant number of tests that are not covered by existing methods. In contrast, Auto-TDD, SWE-Agent+, AutoCodeRover, and Libro each produce fewer unique test cases, ranging from 5 to 6.  
Overall, the results demonstrate that \toolname{} not only complements prior techniques, but also is the only technique to successfully handle a larger number of cases.

\begingroup
\setlength{\abovecaptionskip}{2pt}  
\begin{figure}[t]
    \centering
    \begin{lstlisting}[language=Python, frame=single, basicstyle=\ttfamily\footnotesize]
import pytest
from _pytest._io.saferepr import saferepr

class SomeClass:
    def __getattribute__(self, attr):
        raise Exception("Attribute access failed")

    def __repr__(self):
        raise Exception("Representation failed")

def test_repr_internal_error_handling():
    with pytest.raises(Exception, match="Attribute access failed"):
        SomeClass().attr

def test_repr_internal_error_in_repr():
    obj = SomeClass()
    with pytest.raises(Exception, match="Representation failed"):
        repr(obj)

def test_repr_internal_error_in_repr_handling():
    obj = SomeClass()
    try:
        repr(obj)
    except Exception:
        pass
    with pytest.raises(Exception, match="Representation failed"):
        repr(obj)
    \end{lstlisting}
    \caption{Test case exclusively generated by \toolname{} for \emph{pytest-7168}.}
    \label{fig:pytest-7168-test}    
\end{figure}
\endgroup

\begingroup
\setlength{\abovecaptionskip}{0pt}  
\setlength{\textfloatsep}{6pt plus 1pt minus 1pt}
\begin{figure}[h]
    \centering
    \begin{graybox}
        \textbf{Direct Match:} Yes  

        \textbf{Reason:}  
        The failure is expected due to the unimplemented feature in the GitHub issue, which describes an \texttt{INTERNALERROR} when an exception occurs in the \texttt{\_\_repr\_\_} method.  

        The test case is designed to check the behavior of \texttt{\_\_repr\_\_} when an exception is raised in \texttt{\_\_getattribute\_\_}, which aligns with the issue.
    \end{graybox}
    \caption{Assertion failure reasoning for \emph{pytest-7168}.}
    \label{fig:pytest-7168-assertion-reasoning}
\end{figure}
\endgroup

\subsubsection{Examples of Successful cases}

Figure~\ref{fig:pytest-7168-test} presents a test case exclusively generated by \toolname{}. None of the baseline approaches are able to generate this F→P test case for the corresponding GitHub issue.  
Once the test is executed, \toolname{} processes the failure logs and applies the assertion check logic to assess whether this failure is expected given the issue. The output of this reasoning step is presented in Figure~\ref{fig:pytest-7168-assertion-reasoning}. This shows that the success of \toolname{} is its ability to analyze the assertion failure and correlate it with the issue description. By leveraging its assertion matching mechanism, \toolname{} determines that the test failure aligns with the reported bug.

The following examples illustrate the iterative refinement that \toolname{} uses to generate an F→P test case for \emph{django\_\_django-12284}. In this case, the initially generated test fails to compile. The error categorization step (Line~\ref{alg1-3} in Algorithm~\ref{alg:test-refinement}) identifies it as a compilation failure and produces the output shown in Figure~\ref{fig:django-12284-initial-failure}.  
Then, the repair step (Line~\ref{alg1-6}) is invoked, as the failure does not match with Step 2, Step 3, or Step 4 of the algorithm. Once the revised test is generated in this step, it undergoes execution again (Line~\ref{alg1-1}). Based on the test execution outcome, the test is then re-categorized. Finally, the assertion matching logic (Line~\ref{alg1-5}) determines that the assertion failure corresponds to the reported issue (Figure~\ref{fig:django-12284-assertion-match}).

\begingroup
\setlength{\abovecaptionskip}{2pt}  
\begin{figure}[t]
    \centering
\begin{graybox}

\textbf{Error category:} Compilation Error

\textbf{Explanation:} The test cannot be executed because the model class \code{'B'} is not properly configured within an application in \code{INSTALLED\_APPS}.

\textbf{Root cause:} The model class \code{'B'} does not ...(shortened for paper)

\textbf{Repair steps:} 

1. Add an explicit \code{app\_label} to the class B in the Meta class. 

2. Ensure that \code{'test\_app'} is correctly listed in the \code{INSTALLED\_APPS} configuration.

\end{graybox}
    \caption{Compilation failure detection in the initially generated test for \emph{django-12284}}
    \label{fig:django-12284-initial-failure}
\end{figure}
\endgroup

\begingroup
\setlength{\abovecaptionskip}{2pt}  
\begin{figure}[t]
    \centering
\begin{graybox}

\textbf{Direct match:} Yes

\textbf{Reason:} The assertion failure occurs because the method \code{get\_field\_foo\_display()} does not return the expected output for the new choice 'C', which is directly related to the bug described in the GitHub issue... (shortened)

\end{graybox}
    \caption{Assertion match after refinement step for \emph{django-12284}}
    \label{fig:django-12284-assertion-match}
\end{figure}
\endgroup

\subsubsection{Example of Failure Cases}

In the \emph{sphinx-doc} project, 11 issue instances are analyzed, including 3 from SWE-verified. None of the baseline techniques, including \toolname, can generate F→P tests for this project.  
Of these 11 cases, 5 fail due to runtime errors, while 1 encounter a test collection error caused by an incompatible extension. For 2 instances (\emph{sphinx-doc\_\_sphinx-8801} and \emph{sphinx-doc\_\_sphinx-8506}), we are unable to build the required Docker images. Interestingly, in 3 cases, assertion failures occurred both before and after the patch was applied, preventing a successful fail-to-pass transition.  
One such case is \emph{sphinx-doc\_\_sphinx-8713}, where the generated test fails with an assertion error. Based on LLM prompting, this is treated as the terminating condition, as the test assertion is identified as relevant to the GitHub issue, as shown in Figure~\ref{fig:sphinx-8713}. 


In the \emph{mwaskom} project, there are 4 instances, with 1 successfully generating an F->P test using LIBRO, SWE-Agent+, SWE-Agent, and Auto-TDD. However, \toolname{} is unable to generate any F->P tests for this case. For \emph{mwaskom\_\_seaborn-3407}, which is successfully reproduced by the baselines, the test generated by \toolname{} resulted in an assertion failure both before and after the patch. For the remaining 3 cases, we encounter runtime failures.

\begingroup
\setlength{\abovecaptionskip}{2pt}  
\begin{figure}[t]
    \centering
    \begin{graybox}    
        \textbf{Direct match:} Yes  

        \textbf{Reason:}  
        The assertion failure is due to the missing implementation of the feature described in the GitHub issue, which states that \texttt{napoleon\_use\_param} should also affect the "Other parameters" section... (shortened)
    \end{graybox}
    \caption{Assertion match reasoning for \emph{sphinx-8713}.}
    \label{fig:sphinx-8713}
\end{figure}
\endgroup




\subsection{Influence of Different Components of \toolname{} (RQ2)}  

\begin{figure}[t]
    \centering
    \includegraphics[width=1\linewidth]{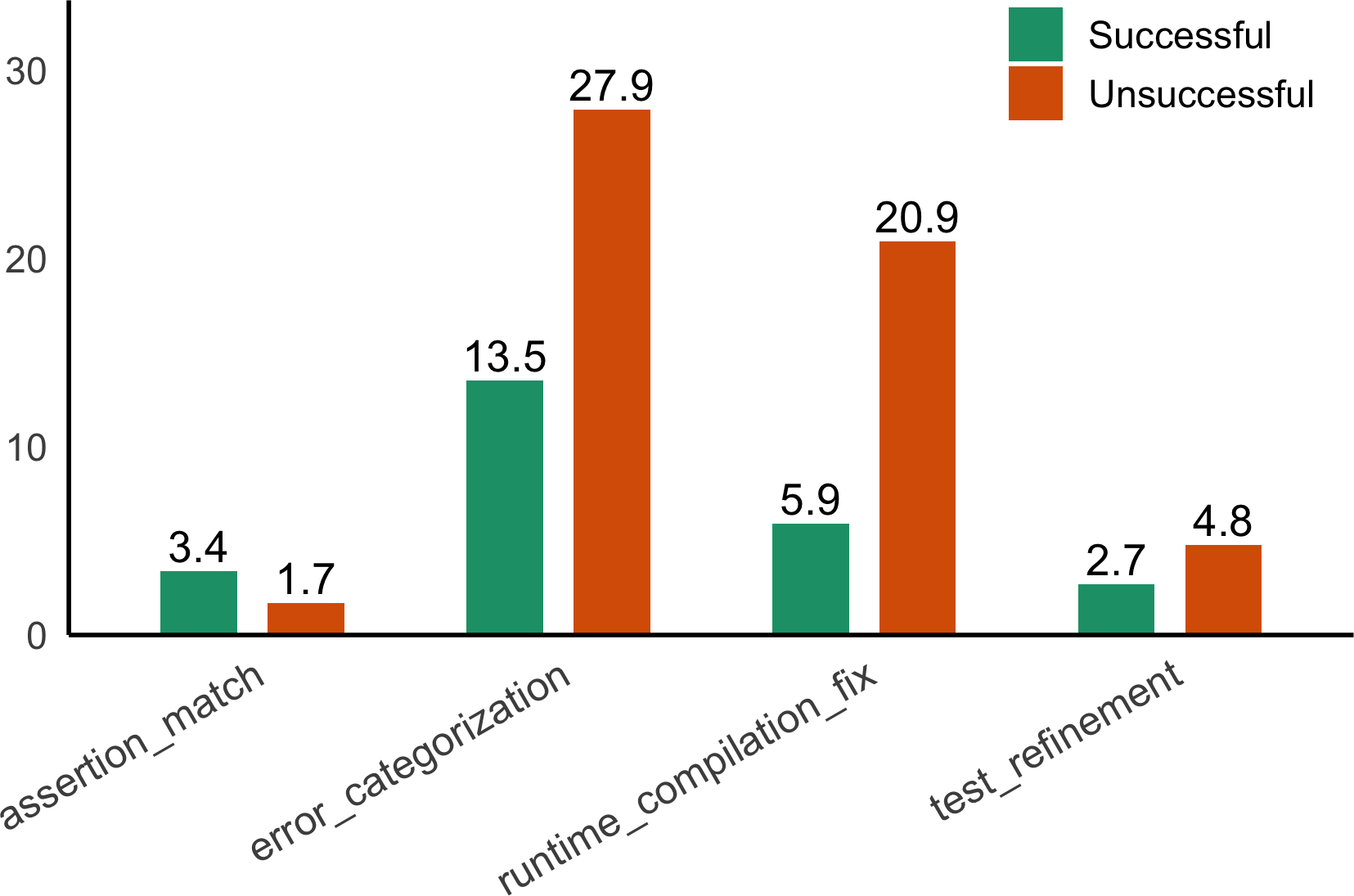}
    \caption{Frequency of invoked steps (average per issue).}
    \label{fig:success_vs_failure}
\vspace{-10pt}
\end{figure}

To understand how different components of the approach contribute to the test generation, we analyze their role in producing successful and unsuccessful test cases. As shown in Figure~\ref{fig:success_vs_failure}, we examine four key components: \textit{assertion matching}, \textit{error categorization}, \textit{runtime and compilation error handling}, and \textit{test refinement}.  

Assertion matching exhibits the highest frequency in successful test cases relative to unsuccessful ones, indicating that when \toolname{} correctly identifies an issue-related assertion, it reliably generates a valid test. However, when the assertion does not align with the issue, additional refinement is required.  

Error categorization plays a crucial role in distinguishing relevant failures from unrelated ones. It is invoked on average 13.5 times per issue, suggesting that repeated classification of test outcomes is central to the refinement loop. Accurate failure classification contributes significantly to effective test generation. 

Runtime and compilation error handling presents a major challenge. It is invoked an average of 20.9 times per unsuccessful issue. This suggests that resolving test cases that crash or fail to compile is a significant bottleneck, limiting the overall effectiveness of \toolname{}.

Test refinement, designed to improve failing test cases, is invoked least often (2.7 times on average for successful cases and 4.8 for unsuccessful ones), suggesting that current refinement strategies may be less effective and could benefit from improved test transformation heuristics.

These results indicate that failure categorization and assertion matching contribute positively to test generation, whereas runtime and compilation errors are the primary barriers to success. Enhancing failure resolution and refinement mechanisms could further improve \toolname{}’s ability to generate high-quality, issue-reproducing tests.

\begin{figure}
    \centering
    \begin{subfigure}[b]{0.4\linewidth}
        \centering
        \includegraphics[width=\linewidth]{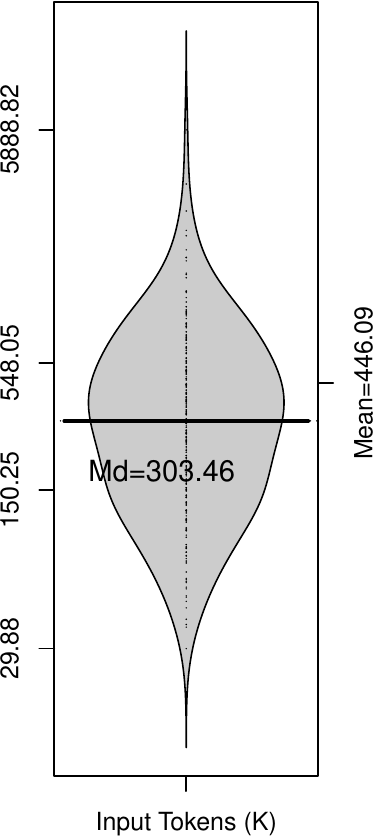}  
    \end{subfigure}
    \hfill
    \begin{subfigure}[b]{0.4\linewidth}
        \centering
        \includegraphics[width=\linewidth]{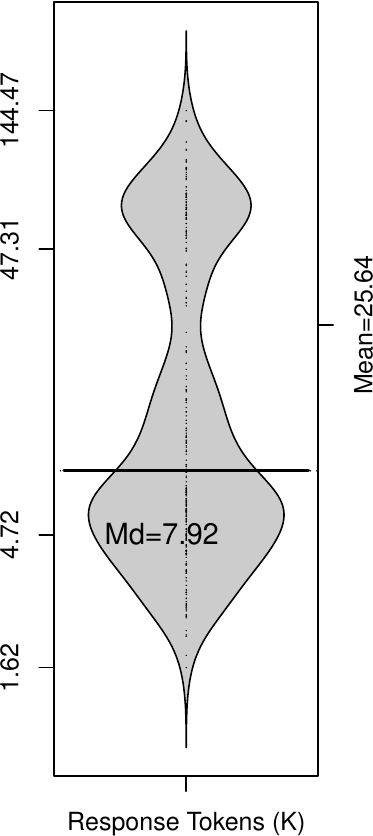}
    \end{subfigure}    
    \caption{Comparison of input tokens and output tokens}
    \label{fig:tokens_comparison}
\end{figure}

\subsection{RQ3: Cost}

Figure~\ref{fig:tokens_comparison} presents the distribution of input and output tokens used by \toolname{}. The median input token consumption is 303.46K, with a mean of 446.09K, whereas the median output token count is significantly lower at 7.92K, with a mean of 25.64K. This disparity arises because \toolname{} invokes the LLM primarily for generating new test code, categorizing errors, and reasoning about failures, all of which require fewer response tokens. In contrast, input token consumption is higher due to the inclusion of GitHub issue descriptions, test case generation guidelines, and execution traces, which extend the context length. 

The cost distribution is shown in Figure~\ref{fig:totalcost} in USD, visualized using a \textit{bean plot}. The median cost per test case generated by \toolname{} is 5.21 cents, with an average cost of 8.23 cents. The total cost per test case ranges from 0.66 to 97 cents. 

Certain cases incur higher costs due to excessive LLM invocations. Specifically, in GitHub projects such as Django, when \toolname{} encounters a runtime error unrelated to the reported issue, it enters an iterative refinement loop, repeatedly invoking the LLM in an attempt to resolve the failure. This leads to substantially higher input token consumption.  

\begingroup
\begin{figure}
    \includegraphics[width=\linewidth]{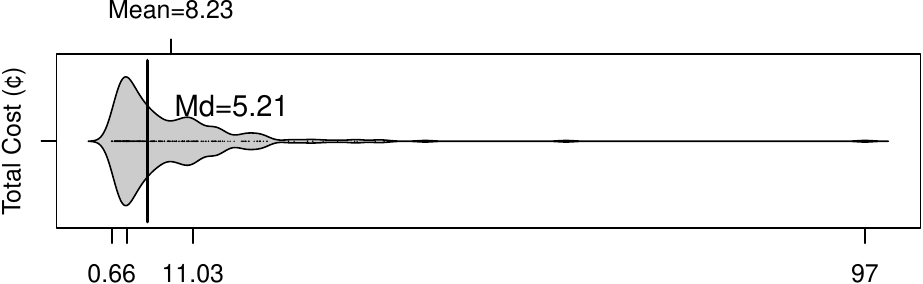}
    \caption{LLM cost in cents (USD) per issue. }
    \label{fig:totalcost}    
\end{figure}
\endgroup

\section{Discussion}
\label{sec:discussion}

\header{Quality of Generated Tests}
Following studies on the quality of generated tests and their potential smells~\cite{10336341, galindo2025increasing}, we manually inspect a sample of successfully generated tests to check for smells and quality issues.
First, we randomly sample, up to three successfully generated tests per project (some projects have fewer than three). This results in a total of 21 tests from 8 projects. Second, we manually go through the generated tests and their corresponding execution traces. For each generated test, we look for the categories of smells described in \cite{10336341}, namely: (1) Act–Assert Mismatch, (2) Redundant Code, (3) Failed Setup, (4) Accessors and Constants. We only find instances of \textit{redundant code} in the inspected sample. Specifically, 5/21 tests had a duplicate setup, i.e., two or more test functions instantiate the same variables with identical values and identical subsequent changes to the variables. Furthermore, 2/21 tests had a duplicate test scenario, i.e, the same test scenario is implemented in two different test cases. The percentage of these two effects is similar to the values reported in prior work~\cite{10336341}. On the positive side, \toolname does not show any smell symptoms from the other three categories.
One reason for the duplication may be that our prompt instructs the LLM to distribute different test scenarios and covered branches over different functions, to isolate points of failure.
Future work could investigate how to guide LLMs toward tests that avoid duplication and other test smells. 


\header{Reproduction Success by Issue Category}
To assess whether \toolname{} depends on the kind of issue addressed, we classified all 276 SWT-bench-lite instances according to the SWE-bench taxonomy~\cite{swebench}. The full set comprises 53 bug fixes (19.2\%), 11 feature requests (4.0\%), no regressions (0.0\%), and 212 other issues (76.8 \%), the latter dominated by maintenance-oriented tags such as ``help wanted''. On the 84 issues for which \toolname{} generated reproducing tests, the distribution is 17 bugs (20.2 \%), 4 features (4.8 \%), 0 regressions, and 63 Others (75.0 \%). These correspond to per-category success rates of 32.1\% for bugs (17/53), 36.4\% for feature requests (4/11), and 29.7\% for other issues (63/212). The maximum gap across category is 6.7\%, indicating that \toolname{} performs consistently on bug fixes, feature requests, and maintenance tasks. Improving reproduction accuracy for maintenance-style (``Other'') issues may require incorporating richer contextual information, such as more detailed task descriptions, to guide the synthesis of targeted and precise test cases.

\header{Threats to Validity}
One potential threat to internal validity is the evaluation dataset size. We evaluate \toolname on SWT-bench-lite, which consists of 276 issues from the larger SWT-bench dataset. While this benchmark has been widely used in prior studies, it may not fully capture the diversity of issue types encountered in practice (e.g, some projects in SWT-bench-lite have less than 10 issues). Another limitation of \toolname is its reliance on meta-prompting, which is particularly beneficial for popular open-source projects. However, while this approach enhances test generation accuracy, it raises concerns about \toolname's ability to generalize to private repositories or lesser-known projects where LLMs lack prior exposure. 

\header{Reproducibility} Our implementation and instructions for reproducing the results are available~\cite{repo}.

\section{Related Work}
\header{Reproducible Test Case Generation}  
\label{sec:reproducible-test-generation}
Reproducing software failures from issue reports is essential for debugging~\cite{beller:debugging:icse18}, regression testing~\cite{libro:icse23}, and automated program repair~\cite{rondon:agent-based-repair-google:arxiv25, multihunk-repair:ase25}. In Section~\ref{sec:baselines}, we discussed approaches for reproducible test case generation. Additionally, BRT Agent~\cite{cheng:agentic-bug-reproduction:arxiv25} focused on generating tests from bug reports, specifically in an industrial setup. The reported success of BRT demonstrates that LLM-based test generation is beneficial not only for public issues but also for private internal issues at large companies such as Google.

\header{Benchmarks for Evaluating Test Generation from GitHub Issues}  
Several benchmarks have been developed to evaluate the effectiveness of automated test generation from GitHub issues. These benchmarks provide real-world issues, developer-written patches, and test cases, enabling systematic assessment of LLM-based test generation techniques. SWE-bench~\cite{swebench} and its derivatives such SWTBench~\cite{niels:bug-fixes-with-code-agents:nips24}, and TDD-Bench-Verified~\cite{autotdd:arxiv24} focus on GitHub issue resolution by pairing reported issues with their corresponding fixes and test cases. Other benchmarks, such as Defects4J~\cite{defects4j}, BugsInPy~\cite{bugsinpy}, and Bugswarm~\cite{bugswarm} provide datasets for evaluating software testing and program repair techniques. While not specifically tailored for test generation from GitHub issues, they are still taken from real fixed issues and provide some context for bug reproduction.


\header{LLM-Based Test Generation} LLMs have been applied to test generation~\cite{feng:bug-replay-llm:icse24, libro:icse23, schafer:test-generation-using-llm:arxiv23, schafer:unit-test-generation-using-llm:tse23, cedar}, with efforts to enhance quality by overcoming coverage plateaus~\cite{lemieux:codamosa:icse23}, incorporating code-aware prompting~\cite{symprompt}, and leveraging coverage information~\cite{coverup:arxiv24, panta}. LLMs have also been explored for augmenting existing tests~\cite{test-improvement-using-llms-meta} refines human-written test suites to improve coverage. Our work differs by focusing not on direct code generation but on synthesizing test cases specifically for reproducing issues raised on GitHub and revealing the buggy behavior related to the issue.

\header{LLMs for GitHub Issue Resolution} LLMs are increasingly being leveraged to automate GitHub issue resolution by generating patches~\cite{magis:neurips25}. Recent research has explored LLM-based agents that analyze issue reports, suggest code modifications, and improve software maintenance workflows. A key area of focus is automated program repair, where agents such as RepairAgent~\cite{repairagent:icse25} apply LLMs to detect and fix defects in source code. Other approaches, including SWE-Agent~\cite{sweagent:arxiv24}, MarsCode Agent~\cite{marscode-agent:arxiv24}, Magis~\cite{magis:neurips25}, and AutoCodeRover~\cite{autocoderover}, tackle broader issue resolution tasks such as bug fixes, feature additions, and code enhancements. Our work, which automatically creates reproducing test cases for said issues, could help these approaches by: i) augmenting existing test suites and thus receiving more feedback about the behavior of code. ii) making the repair approaches more practical and realistic by using automatically generated test cases instead of developer-written tests, often, added after fixing the bug.

\header{Execution-Oriented Agents}
\label{sec:comparison-executionagent}
%
ExecutionAgent~\cite{agent-test-execution:arxiv24} and \toolname{} both use LLMs to automate software testing tasks but differ significantly in goals and techniques. ExecutionAgent focuses on automating the setup and execution of existing test suites by inferring and running project-specific build, dependency, and test commands. It assumes tests already exist and prioritizes execution without human input. In contrast, \toolname{} generates new test cases to reproduce failures described in issue reports, operating under the assumption that no reproducing test exists. It takes issue descriptions and stack traces as input to synthesize failing test logic. This means, while ExecutionAgent focuses on command line actions such as installing dependencies, \toolname{}, in contrast, focuses on generating test code. Subsequently, meta-prompting in ExecutionAgent helps infer commands and actions tied to the repository structure and languages  while \toolname{} uses it to reflect project-specific test idioms such as naming, fixtures, and assertions. Though the principle of meta-prompting is the same, \toolname{} adapts its own meta-prompting for the target task. Furthermore, \toolname{} uses a similarity-based approach to find relevant test and code files while ExecutionAgent just relies on prompting and meta-promting to find relevant files at the start.

In addition, ExecutionAgent uses runtime feedback to fix command execution errors (e.g., build failures), refining its inferred commands. \toolname{} uses runtime diagnostics such as stack traces and assertion errors to iteratively refine the generated tests. It also introduces root cause hypothesis branching to explore alternative failure explanations and generate diverse test cases. This failure reasoning and root cause analysis approach is absent in ExecutionAgent.

On a high-level, while both systems rely on LLMs and meta-prompting, they serve complementary purposes: ExecutionAgent automates test execution, while \toolname{} enables semantic test synthesis from natural language and runtime signals.

\section{Conclusion}
In this paper, we introduce \toolname{}, a novel approach for automated generation of issue-reproducing test cases, addressing a key challenge in software testing and debugging. Our evaluation on the SWT-bench-lite dataset resulted in \toolname{} outperforming the best baselines by an 8.7\% margin in issue reproduction, successfully generating test cases for 28 previously unreproduced issues, and contributing to 68.3\% of the total issues reproduced across all tools.
Beyond issue reproduction, our approach has broader implications for automated debugging, regression testing, and program repair. By bridging the gap between issue descriptions and executable test cases, \toolname{} streamlines the debugging process and facilitates more reliable automated program repair. Future work includes extending our methodology to support a wider range of failure scenarios, improving test adaptation strategies, and integrating Issue2Test with existing automated repair pipelines to further enhance software reliability and maintainability.

\begin{acks}
This work was supported in part by the Canadian Natural Sciences and Engineering Research Council (NSERC DG), Amazon Research Awards (AWS Generative AI), the European Research Council (ERC; grant agreements 851895 and 101155832), and the German Research Foundation (DFG; projects 492507603, 516334526, and 526259073).
\end{acks}

\bibliographystyle{ACM-Reference-Format}
\bibliography{references}

\end{document}